\documentstyle[aps,prl,multicol]{revtex}
\begin{document}
\draft
 \def\OP{\tensor P}
\def\B.#1{{\bbox{#1}}}
\def\BE {\begin{equation}}
\def\EE {\end{equation}}
\def\BEA {\begin{eqnarray}}
\def\EEA {\end{eqnarray}}
\def\Fbox#1{\vskip1ex\hbox to 8.5cm{\hfil\fboxsep0.05cm\fbox{%
  \parbox{8.5cm}{#1}}\hfil}\vskip1ex}

\title{{\rm PHYSICAL REVIEW LETTERS \hfill }
{\sl Submitted 21 May 1997}\\~~\\
Exact Result for the 3rd Order Correlations of Velocity in
  Turbulence with Helicity} \author {Victor S. L'vov$^{1,2}$, Evgenii
  Podivilov$^{1,2}$ and Itamar Procaccia$^1$}
\address{${^1}$Department of~~Chemical Physics,
 The Weizmann Institute
  of Science,
  Rehovot 76100, Israel,\\
  ${^2}$Institute of Automation and Electrometry, Ac. Sci.\ of Russia,
  630090, Novosibirsk, Russia} \maketitle
\begin{abstract}
  All statistical models of turbulence take into account Kolmogorov's
  exact result known as the "4/5 law" which stems from energy
  conservation. This law states that the energy flux expressed as a
  spatial derivative of the 3rd order velocity correlator equals the
  rate of energy dissipation. We have found an additional exact result
  which stems from the conservation of helicity in turbulence without
  inversion symmetry. It equates the flux of helicity expressed as a
  second spatial derivative of the 3rd order velocity correlator with
  the rate of helicity dissipation. This exact result must be
  incorporated to all statistical theories of turbulence with
  helicity.
 After submitting this paper for publication we learned that   
 the main result was independently found by  Otto Chkhetiani in 
  JETP Lett . {\bf 63}, 808 (1996).
\end{abstract}
\begin{multicols}{2}
One of the best known results in the statistical theory of turbulence
is Kolmogorov's ``four-fifth law" which was discovered in 1941
\cite{41Kol}.  This law pertains to the third order moment of
longitudinal velocity differences $\delta u_l(\B.r,\B.R,t)\equiv
[\B.u(\B.r+\B.R,t)-\B.u(\B.r,t)]\cdot\B.R/R$ where $\B.u(\B.r,t)$ is
the Eulerian velocity field of the turbulent fluid. The fourth-fifth
law states that in homogeneous, isotropic and stationary turbulence,
in the limit of vanishing kinematic viscosity $\nu\to 0$
\Fbox{\begin{equation}
\left<[\delta u_l(\B.r,\B.R,t)]^3\right> = -\case{4}{5}
 \bar\epsilon R \ ,
\label{4/5}
\end{equation}}
\noindent
where the symbol $\left<\dots\right>$ stands for an ensemble average
over $\B.r$ and $t$, and $\bar\epsilon$ is the mean energy flux per
unit time and mass $\bar\epsilon\equiv \nu\left<|\nabla_\alpha
  u_\beta|^2\right>$. The only assumption needed to derive this law is
that the dissipation is finite in the limit $\nu\to 0$. As noted by
Frisch, ``this is one of the most important results in fully developed
turbulence because it is both exact and nontrivial. It thus
constitutes a kind of `boundary condition' on theories of turbulence:
such theories, to be acceptable, must either satisfy the four-fifth
law, or explicitly violate the assumptions made in deriving it"
\cite{Fri}.

In this Letter we derive an additional exact relation that appears to
have the same status as the fourth-fifth law, pertaining to
homogeneous, stationary and isotropic turbulence with helicity.
Defining the velocity $\B.v(\B.r,t)$ as $\B.v(\B.r,t)\equiv
\B.u(\B.r,t) -\left<\B.u\right>$ we consider the simultaneous 3rd
order tensor correlation function which depends on two space points:
\begin{equation}
J^{\alpha,\beta\gamma}(\B.R)\equiv \left<v^\alpha
(\B.r+\B.R,t)v^\beta(\B.r,t)
v^\gamma(\B.r,t)\right> \ . \label{defJ}
\end{equation}
We show that in the limit $\nu\to 0$, under the same assumption
leading to the fourth-fifth law, this correlation function reads
\begin{eqnarray}
J^{\alpha,\beta\gamma}(\B.R)=&-&{\bar\epsilon \over
10}(R^\gamma\delta_{\alpha\beta}
+R^\beta\delta_{\alpha\gamma}-{2\over
3}R^\alpha\delta_{\beta\gamma})\nonumber\\ &-&
{\bar h\over 30}
(\epsilon_{\alpha\beta\delta}R^\gamma
+\epsilon_{\alpha\gamma\delta}R^\beta)
R^\delta \ , \label{result}
\end{eqnarray}
where $\delta_{\alpha\beta}$ is the Kronecker delta and
$\epsilon_{\alpha\beta\gamma}$ is the fully antisymmetric tensor. The
quantity $\bar h$ is the mean dissipation of helicity per unit mass
and time,
\begin{equation}
\bar h \equiv \nu \left<(\nabla^\alpha
u^\beta)(\nabla^\alpha[\B.\nabla\times \B.u]^\beta)\right>
\ , \label{defh}
\end{equation}
where repeated indices are summed upon. The new result (\ref{result})
can be also displayed in a form that depends on $\bar h$ alone by
introducing the longitudinal and transverse parts of $\B.u$: the
longitudinal part is $\B.u_l\equiv \B.R(\B.u\cdot \B.R)/R^2$ and the
transverse part is $\B.u_t \equiv \B.u-\B.u_l$.  In addition we have
$\delta\B.u_l(\B.r,\B.R,t)\equiv \delta u_l(\B.r,\B.R,t) \B.R/ R$. In
terms of these quantities we can propose a ``two fifteenth law"
\Fbox{
\begin{equation}
\left<[\delta\B.u_l(\B.r,\B.R,t)]\cdot [\B.u_t(\B.R+\B.r,t)
\times \B.u_t(\B.r,t)]\right> =\case{2}{15}\bar h R^2 \ . \label{short}
\end{equation}
}\noindent
We note that this result holds also when we replace $\B.u$ by $\B.v$
everywhere.

To derive the result (\ref{result}) we start from the correlation
function $J^{\alpha,\beta\gamma}(\B.R)$ which is symmetric with
respect to exchange of the indices $\beta$ and $\gamma$ as is clear
from the definition.  In an isotropic homogeneous medium with helicity
(no inversion symmetry) the relevant symmetry group is the rotation
group O(3) whose irreducible representations can be expressed using
the spherical harmonics $Y_{\ell,m}$.  The most general form of this
object has contributions from $\ell=1,2$ and 3:
\begin{eqnarray}
&&J^{\alpha,\beta\gamma}(\B.R)=a_1(R)[\delta_{\alpha\beta}R^\gamma+
\delta_{\
alpha\gamma}
R^\beta+\delta_{\beta\gamma}R^\alpha] \label{general} \\
&&+\tilde a_1(R)[\delta_{\alpha\beta}R^\gamma+\delta_{\alpha\gamma}
R^\beta-2\delta_{\beta\gamma}R^\alpha]\nonumber \\
&&+b_2(R)[\epsilon_{\alpha\beta\delta}
R^\gamma+\epsilon_{\alpha\gamma\delta} R^\beta]R^\delta
\nonumber \\ &&+a_3(R)[\delta_{\alpha\beta}R^\gamma+
\delta_{\alpha\gamma}
R^\beta+\delta_{\beta\gamma}R^\alpha-5R^\alpha R^\beta 
R^\gamma/R^2] \ .
\nonumber
\end{eqnarray}
The coefficients in this expression carry the $\ell$ index,
multiplying terms that are irreducible representations of the rotation
group of dimension $2\ell+1$.  The dimension of the irreducible
representation is the number of tensor components that transform to
each other upon rotations of the system of coordinates.  All the
tensor components of a given irreducible representation with $\ell=1$,
$\ell=2$ or $\ell=3$ must have the same coefficient which depends only
on $R$. This general representation is invariant to the choice of
orientation of the coordinates.

Not all the coefficients are independent for incompressible flows.
Requiring $\partial J^{\alpha,\beta\gamma}(\B.R)/\partial R^\alpha=0$
leads to two relations among the coefficients:
\begin{eqnarray}
&&\Big({d\over dR}+{5\over R}\Big)a_3(R)={2\over 3}
{d\over dR}\big[a_1(R)
+\tilde a_1(R)\big]  \ , \label{increl} \\
&&\Big({d\over dR}+{3\over R}\Big)\big[5a_1(R)
-4\tilde a_1(R)\big]=0 \ .
\nonumber
\end{eqnarray}
As we have two conditions relating the three coefficients $a_1,\tilde
a_1$ and $a_3$ only one of them is independent. Kolmogorov's
derivation \cite{41Kol} related the rate of energy dissipation to the
value of the remaining unknown. Here the coefficient $b_2$ remains
undetermined by the incompressibility constraint; it will be
determined by the rate of helicity dissipation.

Kolmogorov's derivation can be paraphrased in a simple manner. Begin
with the second order structure function which is related to the
energy of $R$-scale motions
\begin{equation}
S_2(R) \equiv \left<|\B.u(\B.R+\B.r,t)
-\B.u(\B.r,t)|^2\right> \ . \label{S2}
\end{equation}
Computing the rate of change of this (time-independent) function from
the Navier-Stokes equations we find
\begin{equation}
0={\partial S_2(R)\over 2\partial t}=-{\cal
D}_2(R)-2\bar\epsilon+\nu\nabla^2S_2(R)
\ , \label{bal}
\end{equation}
where ${\cal D}_2(R)$ stems from the nonlinear term
$(\B.u\cdot\B.\nabla)\B.u$ and as a result it consists of a
correlation function including a velocity derivative.  The
conservation of energy allows the derivative to be taken outside the
correlation function:
\begin{eqnarray}
{\cal D}_2(R) \equiv &&{\partial\over \partial R^\beta}
\langle u^\alpha(\B.r,t)u^\alpha(\B.r+\B.R,t)
\big[u^\beta(\B.r,t)\label{D2}\\ &&-
\!u^\beta(\B.r+\B.R,t)\big]\rangle \ .
\nonumber
\end{eqnarray}
In terms of the function of Eq. (\ref{defJ}) we can write
\begin{equation}
{\cal D}_2(R) =  {\partial\over \partial
R^\beta}\Big[J^{\alpha,\beta\alpha}(\B.R,t)-
J^{\alpha,\beta\alpha}(-\B.R,t)\Big] \ . \label{relate}
\end{equation}
Note that Eq. (\ref{defJ}) is written in terms of $\B.v$ rather than
$\B.u$, but using the incompressibility constraint we can easily prove
that Eq. (\ref{D2}) can also be identically written in terms of $\B.v$
rather than $\B.u$.  We proceed using Eq. (\ref{general}) in Eq.
(\ref{relate}), and find
\begin{equation}
{\cal D}_2(R) =  2{\partial\over \partial
R^\beta}R^\beta\big[5a_1(R)+2\tilde a_1(R)\big]
\ . \label{D2a1}
\end{equation}
For $R$ in the inertial interval, and for $\nu\to 0$, we can read from
Eq.  (\ref{bal}) ${\cal D}_2(R)=-2\bar\epsilon$ and therefore have the
third relation that is needed to solve all the three unknown
coefficients. A calculation leads to
\begin{equation}
a_1(R)=-2\bar\epsilon/45\ , \quad \tilde a_1=-\bar \epsilon/18\ , \quad
a_3=0\ . \label{avalues}
\end{equation}
The choice of the structure function $S_2(R)$ leaves the coefficient
$b_2(R)$ undetermined, and another correlation function is needed in
order to remedy the situation. Since the helicity is $\B.u\cdot
[\B.\nabla\times \B.u]$, we seek a correlation function which is
related to the helicity of eddies of scale of $R$:
\begin{eqnarray}
&&T_2(R)\equiv \langle \big[\B.u(\B.R+\B.r,t)-\B.u(\B.r,t)\big]
\nonumber \\ &&\cdot \big[\B.\nabla\times
\B.u(\B.r+\B.R,t)-\B.\nabla\times 
\B.u(\B.r,t)\big]\rangle \ . \label{T2}
\end{eqnarray}
The proper choice of this correlation function is the crucial idea of
this Letter. The rest is a straightforward calculation. Using the
Navier-Stokes equations to compute the rate of change of this quantity
we find
\begin{equation}
0={\partial T_2(R)\over 2\partial t}=-G_2(R)-2\bar h
 -\nu\nabla^2T_2(R) \ ,
\label{bal2}
\end{equation}
which is the analog of (\ref{bal}), and where
\begin{eqnarray}
&&G_2(R)=\{\langle\B.u(\B.r,t)\cdot\big[\B.\nabla_R
\times\big[\B.u(\B.r+\B.R
,t)\times
\big[\B.\nabla_R \label{G2}\\ &&\times \B.u(\B.r
+\B.R,t)\big]\big]\big]\rangle\}
+\{{\rm term}~\B.R\to -\B.R\} \ . \nonumber
\end{eqnarray}
The conservation of helicity allows the extraction of two derivatives
outside the correlation functions. The result can be expressed in
terms of our definition (\ref{defJ}):
\begin{equation}
G_2(R)={\partial\over \partial R^\lambda}{\partial\over
 \partial R^\kappa}
\epsilon_{\alpha\lambda\mu}\epsilon_{\mu\beta\nu} 
\epsilon_{\nu\kappa\gamma}
\big[J^{\alpha,\beta\gamma}(\B.R)+J^{\alpha,\beta\gamma}
(-\B.R)\big] \ .
\label{G2J}
\end{equation}
Substituting Eq. (\ref{general}) we find
\begin{equation}
G_2(R)=2{\partial^2\over \partial R^\lambda \partial
R^\kappa}b_2(R)\big[R^\lambda R^\kappa
-\delta_{\lambda\kappa}R^2\big] \ , \label{G2b}
\end{equation}
which is the analog of Eq. (\ref{D2a1}).  Using Eq. (\ref{bal2}) in
the inertial interval in the limit $\nu\to 0$ we find the differential
equation
\begin{equation}
R^2{d^2 b_2(R)\over dR^2}+9R{db_2(R)\over dR} 
+15 b_2(R)=-{\bar h \over 2}
\ . \label{difb2}
\end{equation}
The general solution of this equation is
\begin{equation}
b_2(R) = -\bar h / 30+\alpha_1 R^{-5}+\alpha_2 R^{-3}  \ .
\end{equation}
Requiring finite solutions in the limit $R\to 0$ means that
$\alpha_1=\alpha_2=0$.
Accordingly we end up with Eq. (\ref{result}).

In conclusion, every theory of turbulence with helicity, to be
acceptable, must either satisfy the fourth-fifth and the two-fifteenth
laws, or explicitly violate the assumptions made in deriving them.

\end{multicols}
\end{document}